\documentclass[reprint,amsmath,amssymb,aps,prl,longbibliography]{revtex4-1}

\usepackage{graphicx}
\usepackage{dcolumn}
\usepackage{bm}
\usepackage{todonotes}

\DeclareSymbolFont{Eulerscripteusm10}{U}{eus}{m}{n}
\SetSymbolFont{Eulerscripteusm10}{bold}{U}{eus}{b}{n}
\DeclareMathSymbol{\euW}{\mathord}{Eulerscripteusm10}{"57}
\DeclareMathSymbol{\euD}{\mathord}{Eulerscripteusm10}{"44}

\DeclareMathAlphabet{\mathbxit}{\encodingdefault}{\rmdefault}{bx}{it}   

\newcommand{\be}{\begin{eqnarray}}
\newcommand{\ee}{\end{eqnarray}}
\DeclareMathAlphabet{\pazocal}{OMS}{zplm}{m}{n}   

\newcommand{\Mcal}{\pazocal{M}}

\newcommand{\Tcal}{\pazocal{T}}

\newcommand{\ra}{\rangle}
\newcommand{\la}{\langle}

\newcommand*{\at}{@}

\newcommand{\nn}{\nonumber}
\def\wh{\widehat}
\def\dg{\dagger}
\def\df{\overset{\rm df}{=}}
\newcommand{\ket}[1]{\mathop{|#1\rangle}\nolimits}

\newcommand{\kbr}[2]{| #1\rangle\!\langle #2 |}

\newcommand{\adg}{{a^\dg}}         
\newcommand{\Tr}[1]{\mathop{{\mathrm{Tr}}_{#1}}}

\def\b{\beta}

\def\om{\omega}

\def\vt{\vartheta}

\def\om{\omega}

\begin{document}


\title{One-shot decoupling and Page curves from a dynamical model for black hole evaporation}\

\author{Kamil Br\'adler}
\email{kbradler\at uottawa.ca}
\affiliation{
     Department of Mathematics and Statistics, University of Ottawa, Ottawa, Canada
    }
\affiliation{
     Max Planck Centre for Extreme and Quantum Photonics, University of Ottawa, Ottawa, Canada
    }
\affiliation{
    Department of Astronomy and Physics,
    Saint Mary's University,
    Halifax, Canada
    }

\author{Christoph Adami}
\email{adami\at msu.edu}
\affiliation{Department of Physics and Astronomy, Michigan State University, East Lansing, MI 48824}

\begin{abstract}
One-shot decoupling is a powerful primitive in quantum information theory and was hypothesized to play a role in the black hole information paradox. We study black hole dynamics modeled by a trilinear Hamiltonian whose semiclassical limit gives rise to Hawking radiation. An explicit numerical calculation of the discretized path integral of the $S$-matrix shows that decoupling is exact in the continuous limit, implying that quantum information is perfectly transferred from the black hole to radiation. A striking consequence of decoupling is the emergence of an output radiation entropy profile that follows Page's prediction. We argue that information transfer and the emergence of Page curves is a robust feature of any multi-linear interaction Hamiltonian with a bounded spectrum.
\end{abstract}

\pacs{Valid PACS appear here}
\maketitle

The explicit mechanism of black hole unitary evaporation is a major open problem in high-energy physics~\cite{FabbriNavarro2005}. The evaporation of a macroscopic black hole is believed to be well described by a semiclassical result of Hawking~\cite{Hawking1976b}, but it has also been argued that for a sufficiently old black hole the in-falling Hawking quanta (hereafter called the ``partner modes'', as opposed to the radiated modes that constitute the Hawking radiation) must somehow resurface in order to preserve the manifest unitarity of the black hole evolution. Page~\cite{page1993information} argued for the latter using a description based entirely on calculating the entropy of the Hawking radiation as a subsystem of the joint pure state of a black hole and its radiation field, without proposing an explicit interaction that would enable this. In the modern language of quantum information theory, Page's calculation implies that when the black hole entropy vanishes, it is {\em decoupled} from its surroundings. Here we present a model in which black hole modes interact with radiation modes both inside and outside of the horizon in a fully unitary manner, and that allows us to explicitly calculate the entropy of the black hole over time as it gradually decouples from its surroundings.

The modern approach to the resolution of the problem started with Ref.~\cite{hayden2007black}. These authors proposed an abstract mechanism for black hole evaporation based on the recently developed ``decoupling'' framework (see a general version in~\cite{dupuis2014one}). Decoupling theory is a fundamental technique to prove a plethora of quantum communication protocols such as state-merging~\cite{horodecki2007quantum}, the fully quantum Slepian-Wolf (FQSW) protocol~\cite{abeyesinghe2009mother} or the channel capacity coding theorem~\cite{hayden2008decoupling}.

The central result of the one-shot decoupling theorem~\cite{dupuis2010decoupling} implies the existence of a completely positive (CP) map $\wh\Tcal_{A\to B}$ and a state $\rho_{RA}$ where the following condition is satisfied
\begin{equation}\label{eq:decoupling}
\|\wh\Tcal_{A\to B}(U\rho_{RA}U^\dg)-\vt_B\otimes\rho_R\|_1\to0
\end{equation}
for almost all unitaries $U$. Here, $\|A\|_1\df\Tr{}[(A^\dg A)^{1/2}]$ is the trace norm, $R$ is an auxiliary system, $\rho_R=\Tr{A}[\rho_{RA}]$ and $\vt_B$ is a state specified later.
For our purposes it will turn out to be sufficient to consider $\rho_{RA}$ as a maximally entangled state (denoted $\Phi_{RA}$).  Then, decoupling Eq.~(\ref{eq:decoupling}) is equivalent
to the conditions of the entanglement of the $A$ subsystem of $\Phi_{RA}$ being transferred to~$E$ via another CP map $\Tcal_{A\to E}$  called a \emph{complementary} channel to $\wh\Tcal_{A\to B}$.

In what follows we examine decoupling induced by a boson trilinear operator describing the interaction of a quantized black hole's modes with those of radiation inside and outside of the event horizon, in the special case when $\Tcal_{A\to E}$ is an identity,
something that, crucially, is not an assumption of the model but rather an unexpected and natural consequence of the interaction Hamiltonian we use.

In the semiclassical description, the outgoing Hawking radiation shrinks the black hole but the exact mechanism of how the black hole loses its mass over time is beyond the scope of Hawking's original derivation, as in that approach states from past infinity are mapped to future infinity via a Bogoliubov transformation. In order to restore an explicit time-dependence to the evaporation process, we can use the by now well-known insight~\cite{Leonhardt2010,Nationetal2012} (see also~\cite{LewisRiesenfeld1969,Pedrosa1989}) that any solution to a time-dependent quantum harmonic oscillator can be understood in terms of a Bogoliubov transformation that connects the initial to the final frequency, and in particular can be implemented via a time-dependent Hamiltonian of the ``squeezing" type in which the coupling strength $r(t)$ plays the role of the classical time-dependent driving field (we set $\hbar=1$)
\be
H=ir(t)(b^\dg c^\dg- bc\big)\;. \label{class}
\ee
Here, the annihilation operators $b$ and $c$ annihilate radiation modes within and outside of the event (or apparent) horizon, respectively.
Such a description can be used to derive Hawking radiation from a collapsing shell of matter, for example~\cite{Alberghietal2001,Vachaspatietal2007,SainiStojkovic2015}. However, in such a time-dependent semi-classical description, the black hole degrees of freedom are not explicitly quantized and the backreaction of the radiation on the black hole cannot be followed.

 The dynamic model for the evaporation of a black hole that we analyze here can be seen as an an extension of Hawking's semi-classical approach~\cite{hawking1975particle} where the black hole internal degrees of freedom are explicitly quantized to allow us to track the black hole quantum entropy over time. This model is explicitly {\em ad hoc}: the quantum interaction is not derived from a quantum field theory that couples gravitational modes to radiation modes. Rather, it is an extension of the semi-classical Hamiltonian (\ref{class}), and reduces to it when the quantum black hole modes are replaced by their classical expectation values. In the static  infinite-time limit, the quantum Hamiltonian (shown below) therefore implements the standard Bogoliubov transformation between past and future infinity, and reproduces the celebrated Hawking results as has been shown previously~\cite{NationBlencowe2010,Nationetal2012,alsing2015parametric}.

The Hamiltonian we study is the so-called boson trilinear Hamiltonian
\begin{equation}\label{eq:triHam}
  H_{\mathrm{tri}}=ir\big(ab^\dg c^\dg-\adg bc\big)
\end{equation}
often encountered in quantum optics and condensed matter physics~\cite{walls1970quantum,brif1996coherent,agrawal1974dynamics} or quantum optomechanics~\cite{aspelmeyer2014cavity}. In quantum optics, the operators $b$ and $c$ refer to the signal and idler modes, respectively, while $a$ refers to the pump mode. The intriguing formal analogy between black hole radiance and parametric amplification has been noted before~\cite{Gerlach1976,NationBlencowe2010,alsing2015parametric}. As a model of black hole and radiation interactions, the operator $a$ in~ (\ref{eq:triHam}) annihilates black hole modes, whereas $c$ and $b$ are defined as in (\ref{class}).

The black hole mass is a function of the input state $n_a=\la a^\dagger a \ra$ as well as of the coupling strength $r$ (called the {\em gain} or squeezing parameter in quantum optics). In the semi-classical limit, the relationship between $r(t)$ in (\ref{class}) and black hole parameters is simply (in units where $\hbar=c=G=k=1$)~\cite{NationBlencowe2010}
\be
T(t)=\frac{\omega}{2\ln[\tanh^{-1}(r(t)t)]}
\ee
where $T(t)$ is the time-dependent Hawking temperature $T(t)=(8\pi M(t))^{-1}$ and $M(t)$ is the black hole mass.

The initial black hole state can be pure or mixed. For simplicity, we restrict our analysis here to the pure state scenario and a black hole in initial state $\ket{n}_a$ ($n$ modes), but our results naturally carry over to an arbitrary superposition and any mixed initial black hole state with corresponding nonzero initial entropy as we discuss below. We now define the time-evolution operator on a quantum state $|\Psi(t_1)\ra$ as $\ket{\Psi(t_2)}=U(t_2,t_1)\ket{\Psi(t_1)}$ where
\begin{equation}
U(t_2,t_1)=\mathsf{T}e^{-i \int_{t_1}^{t_2} H(t') dt'}.
\end{equation}
The symbol $\mathsf{T}$ stands for Dyson's time-ordering operator and $H(t)$ is the Hamiltonian describing the unitary evolution of the quantum state.
We can then write the time-evolution of the black hole initial state $|\Psi(0)\ra=|n\ra_a|0\ra_b|0\ra_c\equiv |n\ra_a|0\ra_{bc}$ as
\begin{equation}\label{eq:psi}
\ket{\Psi(t)}=U(t,0)\ket{\Psi(0)}=\mathsf{T}e^{-i\int_0^tH_{\mathrm{tri}}(t')dt'}\ket{n}_a\ket{0}_{bc},
\end{equation}
where $H_{\mathrm{tri}}$ is the trilinear Hamiltonian (written as a sum over modes)
\begin{equation}
H_{\mathrm{tri}}=\sum_{k=-\infty}^\infty i r_{\om_k}(t)\big(a_kb_k^\dg c_k^\dg-a_k^\dg b_kc_k\big).
\end{equation}
In the following, we will focus on a single mode $k$ with energy $\omega_k$, and omit the index $k$ for convenience.

To calculate (\ref{eq:psi}), we introduce small time slices $\Delta t$ and discretize the path integral so that with $t=N\Delta t$
\begin{equation}\label{eq:prod}
U(t)=\mathsf{T}e^{-i\int_0^tH_{\mathrm{tri}}(t')}dt'\approx \prod_{i=1}^N e^{-i\Delta t H_i},
\end{equation}
where the $i$-th time-slice Hamiltonian $H_i=ir_0\big(a b_i^\dg c_i^\dg-a^\dg b_ic_i\big)$ acts on the black hole state and the $i$-th slice of the $bc$ Hilbert space $\ket{0}_{b_ic_i}$. The initial value $r_0$ sets the energy scale, and we simply set $r_0=1$ in the following.

We now apply (\ref{eq:prod}), so that
\begin{equation}
\ket{\Psi(t)}_{a'bc}=\euW\ket{n}_a\ket{0}_{bc}=\prod_{i=1}^N e^{-i\Delta t H_i} \ket{n}_a\ket{0}_{bc},
\end{equation}
where $\ket{0}_{bc}\df\ket{0}_{b_Nc_N}\otimes\hdots\otimes\ket{0}_{b_1c_1}$, $W^{(i)}=e^{-i\Delta t H_i}$ and $\euW=W^{(N)}\circ\hdots\circ W^{(1)}$. The first time slice is
\begin{equation}\label{eq:W1action}
\ket{\Psi(1)}=W^{(1)}|n\ra_a\ket{0}_{b_1c_1}=\sum_{j=0}^nU_{nj}^{(1)}\ket{j\ra_a|n-j}_{b_1c_1}
\end{equation}
while $|\Psi(t)\rangle$ is given by
\begin{align}\label{eq:WNaction}
 \ket{\Psi(t)}_{a'bc} & = \sum_{j_1...j_N}U_{nj}^{(1)}\hdots U_{j_{N-1}j_N}^{(N)} \nn\\
 &\times |j_N\rangle_a|j_{N-1}-j_N\rangle_{b_Nc_N}\hdots|n-j_1\rangle_{b_1c_1}.
\end{align}
Note that approximating the path integral using a single time-slice---the static path approximation (SPA), see e.g.~\cite{Arveetal1988,Langetal1993}---can yield good results at very low temperatures, but disregards self-consistent temporal fluctuations. Indeed, SPA calculations of the black hole entropy using the trilinear Hamiltonian lead to an oscillating behavior of the black hole entropy~\cite{NationBlencowe2010,alsing2015parametric}.
\begin{figure}[ht]
   \centering
   \includegraphics[width=8cm]{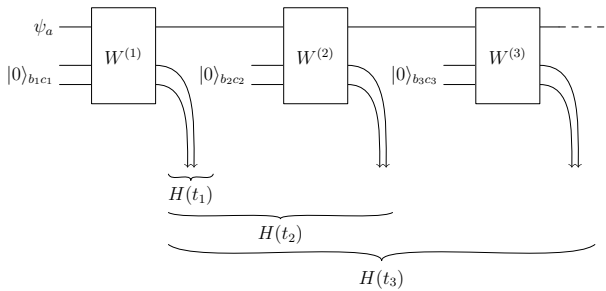}
   \caption{The action of the discretized path integral $W^{(i)}$ is sketched. The braces indicate the entropy calculation of the radiated modes at each time-slice.}
   \label{fig:scheme1}
\end{figure}
The black hole von Neumann entropy can be calculated using the marginal density matrix $\vt_{a'}(t)=\Tr{bc}\big[\kbr{\Psi(t)}{\Psi(t)}_{a'bc}\big]$, which can be written entirely in terms of the coefficients $|U_{ij}|^2$, which describe the probability to observe $j$ modes of the black hole given that in the previous iteration we found $i$ modes) $\vt_{a'}(t)=\sum_{j_N=0}^n\limits p_{j_N}\kbr{j_N}{j_N}$ where
\begin{equation}\label{eq:pi}
p_{j_N}=\sum_{j_1>j_2>\hdots>j_{N-1}}^n |U_{nj_1}^{(1)}|^2\hdots |U^{(N-1)}_{j_{N-2}j_{N-1}}|^2 |U_{j_{N-1}j_N}^{(N)}|^2.
\end{equation}

To follow the time evolution of the black hole entropy $H_n(t)=-\Tr{}\big[\vt_{a'}(t)\log\vt_{a'}(t)\big]$ given the initial state of the black hole is a pure state with $n$ quanta, we need to evaluate the $n$ coefficients $p_{j_N}$ above. This is a difficult calculation because unlike the Hawking isometry, the unitary operator $W^{(i)}$ is not known to be factorizable in a simple way. The usual formal factorization formulas~\cite{magnus1954exponential,suzuki1976generalized,trotter1959product} are infinite operator products and often are not suitable for practical calculations. A common route to obtain the time evolution is to expand the exponential around $r_0t\approx0$ but this allows only a short-time analysis of the evolution operator. For greater $r_0t$, the Taylor expansion is prohibitively inefficient. However, a method developed in Ref.~\cite{Bradler2015} enables us to find the action of the unitary operators even for large values of $r_0t$. The action of $W^{(i)}$  acquires an interesting combinatorial interpretation in terms of an integer lattice known as a generalized Dyck path~\cite{stanley1999enumerative} if it acts on any state generated by the repeated action of $ab^\dg c^\dg$ on a ground state $\ket{\mathrm{gr}}$, defined by $\adg bc\ket{\mathrm{gr}}=0$. The basis elements $\{\ket{k\,00}_{abc}\}_{k=0}^n$ spanning our input Hilbert space of $W^{(i)}$ are all ground states of $H_{i}$.

\begin{figure}[t]
  \includegraphics[width=8cm]{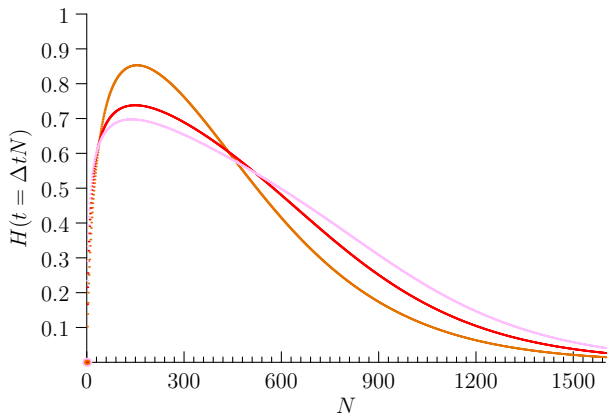}
   \caption{Black hole entropy $H_n(t)=H[\vt_{a'}]=H[\vt_{bc}(t)]$ as a function of the iteration number $N$ for initial states $\ket{n}_a\ket{0}_{bc}$ with $n=5$ (orange), $n=20$ (red) and $n=50$ (pink), with $t=\Delta t N$. Logarithms are base ${n+1}$ so that each black hole entropy $H_n(t)\leq1$.   We use $r_0=1$ and $\Delta t=1/15$. The curves are qualitatively unchanged if $\Delta t=1/25$ and using a commensurately larger number of time slices (data not shown).}
   \label{fig:entropy}
\end{figure}

We evaluated the output entropy for black holes with initial quanta $n$ for $0\leq n\leq50$ using the discretized path integral (\ref{eq:WNaction}). The resulting entropy as a function of the number of time slices (at constant small $\Delta t$) is shown in Fig.~\ref{fig:entropy} for $n=5,20$ and $50$. An entropy curve strikingly similar to the one predicted by Page~\cite{page1993information} emerges in the limit of large $t$ (several thousand time slices). This shape is consequence of the fact that the final black hole density matrix $\vt_{a'}$ (for $N\gg0$) is extremely close to a pure vacuum state $\ket{0}_{a'}$ for all examined input basis states $\ket{n}$. Note that the case $n=50$ gives rise to an extremely large Hilbert space, and to obtain a reliable Taylor expansion for $\Delta t=1/15$ would require expansion to order up to 500. A straightforward numerical evaluation of the $2^{500}$ terms is of course impossible, but the method developed in~\cite{Bradler2015} makes the calculation tractable on a High-Performance Computing Cluster. We have also performed a calculation of the integral where partner modes are allowed to interact with the black hole a second time, breaking the symmetry between partner and Hawking modes. Such a calculation is far more complex, but does not change the shape of the Page curves appreciably.

It is important to stress that even though we studied  the effect of $\euW$ only on basis states $\ket{n}$, we can extend this conclusion to an arbitrary input pure state $\psi_a=\sum_{k=0}^n \b_k\ket{k}_a$. It is sufficient to show decoupling of the spanning basis set. Thus, ultimately the action of $\euW$ can be reformulated as inducing
\begin{equation}\label{eq:arbitraryPurestate}
  \ket{k}_a\ket{0}_{bc}\overset{\euW}{\to} \ket{\Psi(t)}_{a'bc}\approx\ket{0}_{a'}\otimes\,\vartheta_{bc}(\ket{k}),
\end{equation}
where $0\leq k\leq n$ and $\vartheta_{bc}(\ket{k})\df\Tr{a'}\big[\kbr{\Psi(t)}{\Psi(t)}_{a'bc}\big]$. Hence $\vartheta_{bc}(\ket{k})$ is necessarily pure and mutually orthogonal for all $k$. Since $\ket{0}_a$ is a ``constant'' state (independent of $k$), it immediately follows that an arbitrary superposition $\psi_a$ will be transferred to the Hawking modes and their partners ($bc$). These are precisely the circumstances captured by the decoupling theorem in Eq.~(\ref{eq:decoupling}) upon identifying $A\leftrightharpoons a,E\leftrightharpoons bc$ and $B\leftrightharpoons a'$. The presence of an input bipartite state $\Phi_{RA}$ instead of $\psi_A$ is just a convenient reformulation so that if the decoupling (transfer) of $\Phi_{RA}$  is successful, it can be used to perfectly teleport any $\psi_A$.

As is implied by (\ref{eq:arbitraryPurestate}), the black hole evolution ultimately turns out to be an \emph{erasure} map $\wh\Tcal_{a\to a'}:\psi_a\to\Tr{}[\psi_{a}]\kbr{0}{0}_{a'}$, which we interpret as complete evaporation. This immediately implies that the map $\Tcal_{a\to bc}$ must be the identity as announced earlier.  The decoupling theorem then implies~\cite{dupuis2014one} that the input state (and so the quantum information it carries) has been nearly perfectly transferred to the $bc$ Hilbert space. What we found therefore is an explicit (and simple) realization of the one-shot state merging protocol~\cite{berta2009single,dupuis2014one} (generalizing Ref.~\cite{horodecki2007quantum} studied in the asymptotic limit) and thus the action of $\wh\Tcal_{a\to a'}$ in~(\ref{eq:decoupling}) turns out to be a simple case of the FQSW protocol~\cite{abeyesinghe2009mother} (the erasure map is essentially the trace map which is a special case of the partial trace appearing in FQSW).

The question of how exactly the information is encoded in the outgoing radiation described by the density matrix $\vartheta_{bc}$ is non-trivial. 
The mapping $\Tcal_{a\to bc}$ is a highly convoluted identity, something that is expected from the decoupling mechanism. Hence the information is not readily available for a read-out---the state occupying the $bc$ Hilbert space is \emph{scrambled}, or, in quantum information theory jargon, it is a highly entangled quantum code.

The output state is highly entangled and so an observer having access only to the $b$ or $c$ mode cannot learn anything. This is a direct consequence of the symmetry between the modes $b$ and $c$ in the boson trilinear Hamiltonian given by Eq.~(\ref{eq:triHam}), also seen in the action of the unitary $W^{(1)}$ and $\euW$ in Eqs.~(\ref{eq:W1action}) and (\ref{eq:WNaction}), respectively. This is a textbook example of how the impossibility of perfect cloning prevents an observer from obtaining any information from $\vt_b$ or $\vt_c$ alone. If there existed a completely positive map that recovers $\psi_a$ from  $\vt_b$ only, the same operation could recover it from $\vt_c$. But that would result in two nearly perfect copies of $\psi_a$ which is prohibited by quantum mechanics. Using the language of quantum Shannon theory, the quantum capacity of the channel $\Mcal_{a\to b(c)}$ must vanish~\cite{bradler2013capacity}.

The trilinear Hamiltonian Eq.~(\ref{eq:triHam}) that allowed us to carry through this analysis is, to put it bluntly, a guess. It does not follow from an interacting quantum field theory of gravity (such as a locally-flat conformal field theory coupled to a dilaton~\cite{Callanetal1992} (see~\cite{Grumiller2002} for additional references). However, it is manifestly an elegant extension of the semi-classical approach, and it is difficult to imagine that a consistent renormalizable quantum field theory of gravity would not, in an effective limit, reduce to precisely an interaction as we describe. In fact, the mechanism of information conservation implied by our model is an example of
 a much larger class of more complicated Hamiltonians with the proper semiclassical limit
$V_{\om_i}\ket{\mathrm{vac}}=1/{\cosh{r_{{\om_i}}}}\sum_{m=0}^{\infty}\tanh^{m}{r_{{\om_i}}}\ket{m}_{b_i}\ket{m}_{c_i}$.
Indeed, it may be possible to use decoupling as a constraint to impose
conditions on the form of the Bogoliubov transformation that connects the
in-vacuum with the out-vacuum. It is clear from general
arguments~\cite{hossenfelder2015disentangling} that at a minimum extra degrees of
freedom are needed that are coupled to the Hawking modes, and in our model
these degrees of freedom are the quantized black hole modes $a$.

\begin{figure}[h]
  \includegraphics[width=8cm]{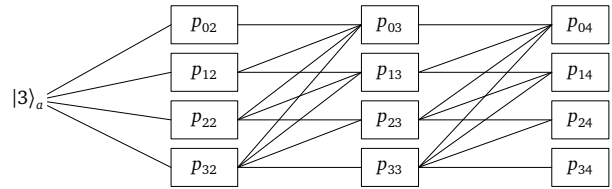}
   \caption{The schematic evolution of the probability Eq.~(\ref{eq:pi}) of the black hole density matrix $\vt_{a'}(t)$ for the input state $\ket{3}_a$. Time goes from left, the boxes are Fock bases ($\ket{0},\dots,\ket{3}$, top to bottom) and the lines represent the contributions from the previous iterations. The vacuum gets the majority of contributions in every iteration (three steps shown here) capturing the black hole evaporation.}
   \label{fig:drift}
\end{figure}

To illustrate the mechanism concretely, we set $\psi_a=\ket{3}_a$ and observe the behavior of the probabilities given by Eq.~(\ref{eq:pi}) in Fig.~\ref{fig:drift}.
The black hole state slowly ``drifts'' towards the vacuum state (the top line) even if the transition amplitudes for the highest Fock state are close to one, that is, the least favorable conditions for the transition to the vacuum state. This is a direct consequence of the spectrum boundedness of the trilinear Hamiltonian Eq.~(\ref{eq:triHam}) -- the number of bosons $n$ in the $a$ mode  that $W=\exp{[-iH_{\rm tri}]}$ can ``distribute'' to the $bc$ modes is naturally equal to or lower than $n$. Crucially, any spectrum-bounded and possibly multilinear Hamiltonian displays exactly the same behavior, and only the information rate differs. As a consequence, the information transfer mechanism--and therefore the Page curve--are expected to remain virtually unchanged even for much more complicated internal dynamics. In other words, the Page curve is robust and universal.

The general features of information transfer between quantum subsystems as described by Page~\cite{page1993information} (see also more recently~\cite{hayden2007black,GiddingsShi2012,MathurPlumberg2011}) have been discussed intently in the past. We believe our calculation is the first explicit realization of these previous abstract arguments, pointing towards decoupling as a mechanism for the resolution of the black hole information loss problem.
Finally, we would like to point out that the explicitly time-dependent black hole evaporation process we described throws new light on the discussion of putative firewalls~\cite{almheiri2013black,braunstein2013better} for old black holes. The usual line of reasoning for firewalls based on the monogamy property of entanglement does not apply here since the black hole state $ \ket{\Psi(t)}_{a'bc}$ from~(\ref{eq:WNaction}) is tripartite entangled.

\begin{acknowledgments}
CA thanks Steve Giddings for discussions. This work was supported in part by Michigan State University through computational resources provided by the Institute for Cyber-Enabled Research.
\end{acknowledgments}

\bibliography{page}

\end{document}